\documentclass{emulateapj}
\usepackage{graphicx,color}

\newcommand{\flux}{erg~s$^{-1}$~cm$^{-2}$~}
\newcommand{\bc}{}

\shorttitle{UV driven evaporation of planets}
\shortauthors{Owen, J. E. \& Alvarez, M. A.}

\begin{document}


\title{UV driven evaporation of close-in planets: energy-limited; recombination-limited and photon-limited flows.}


\author{James E. Owen\altaffilmark{1}}
\affil{Institute for Advanced Study, Einstein Drive, Princeton NJ, 08540, USA}
\email{jowen@ias.edu}
\author{Marcelo A. Alvarez}
\affil{Canadian Institute for Theoretical Astrophysics, 60 St George Street, Toronto, M5S 3H8, ON, CANADA}

%
%


\altaffiltext{1}{Hubble Fellow}


\begin{abstract}
We have investigated the evaporation of close-in exoplanets irradiated by ionizing photons. We find that the properties of the flow are controlled by the ratio of the recombination time to the flow time-scale. When the recombination time-scale is short compared to the flow time-scale the the flow is in approximate local ionization equilibrium with a thin ionization front, where the photon mean free path is short compared to flow scale. In this ``recombination limited'' flow the mass-loss scales roughly with the square root of the incident flux. When the recombination time is long compared to the flow time-scale the ionization front becomes thick and encompasses the entire flow, with the mass-loss rate scaling linearly with flux. If the planet's potential is deep the flow is approximately ``energy-limited''; however, if the planet's potential is shallow we identify a new limiting mass-loss regime, which we term ``photon-limited''. In this scenario the mass-loss rate is purely limited by the incoming flux of ionizing photons. We have developed a new numerical approach that takes into account the frequency dependence of the incoming ionizing spectrum and performed a large suite of 1D simulations to characterise UV driven mass-loss around low mass planets. We find the flow is ``recombination-limited'' at high fluxes but becomes ``energy-limited'' at low fluxes; however, the transition is broad occurring over several order of magnitude in flux. Finally, we point out the transitions between the different flow types does not occur at a single flux value, but depends on the planet's properties, with higher mass planets becoming ``energy-limited'' at lower fluxes.  
\end{abstract}





\section{Introduction}
Observational campaigns to detect exoplanets have been incredibly successful, having found thousands of possible exoplanets (e.g. Batalha et al. 2013). However, what's surprising is that many of the exoplanets are found to be incredibly close to their parent stars with separations $<0.1$~AU being common. Furthermore, while the original detected planets at small separations were massive (of order a Jupiter mass and higher) - so called ``hot jupiters'' - these are relatively rare (Fressin et al. 2013). With the advent of the {\it Kepler} mission the majority of planets found close to their parent star are small radius, low mass objects, with the majority of stars containing at least one small ($\lesssim 4$~R$_\oplus$) planet with a period shorter than Mercury's (Fressin et al. 2013). This indicates that the formation of close-in small planets is a dominant mode of planet formation. 

At such close separations, the environment is very different to planets in our own solar system and irradiation from the central star can play a very important role, particularly when the star is young (e.g. Owen \& Wu, 2013) or during the late stages of stellar evolution (e.g. Bear \& Soker, 2011; 2015). The star's bolometric luminosity can heat-up the planet's atmosphere leading to an increased scale height, as well as preventing the planet from cooling (Guillot et al. 1996, Burrows et al. 2000, Baraffe et al. 2003). Furthermore, the high energy ionizing luminosity from the star can heat-up the upper layers to temperatures of order the escape temperature, leading to mass-loss (e.g. Lammer et al. 2003, Tian et al. 2005, Owen \& Jackson, 2012). 

When the star is relatively young ($<$ several 100~Myr old) the ionizing luminosity can represent a significant fraction  of the total output of the star $\sim 10^{-3}\,L_*$ (Jackson et al. 2012). Integrating the total received  ionizing luminosity over Gyr time-scales and comparing it to the binding energy of close in planets it was realised that planetary evaporation could play a significant evolutionary role and actually remove a large fraction of the planet's atmosphere (Lammer et al. 2003, Jackson et al. 2012), especially at low planet masses (Owen \& Jackson 2012, Lopez et al. 2012, Owen \& Wu 2013, Lopez \& Fortney 2013). 

Due to the importance of the evolutionary effects of planetary evaporation one requires accurate mass-loss rates. Commonly the approach is to turn a certain fraction of the received ionizing luminosity into work/heating and associated mass-loss, either globally (Lammer et al. 2003, Erkaev et al. 2007) and obtain the well known ``energy-limited'' mass-loss rate; or locally in simplified simulations (e.g. Yelle et al. 2004, Tian et al. 2005, Lammer et al. 2013). Such an approach implicitly assumes that $P{\rm d}V$ work dominates the energy loss and radiative cooling or ionization physics are unimportant.

More detailed studies have since been performed that take into consideration some of the details of radiative transfer and ionization chemistry (Koskinen et al. 2007, 2010, 2014; Chadney et al. 2014); however, these models typically concentrate on a handful of observed hot jupiters.  Owen \& Jackson (2012) considered the role X-rays play and showed heating from secondary ionization and line cooling were important when the sonic point occurred in an area of the flow dominated by X-ray heating and argued that in this case the flow was never ``energy-limited''. Owen \& Jackson (2012) also presented a schematic picture of when UV or X-ray heating dominates and argued that X-rays would dominant for the highest fluxes around low-mass planets. 

Furthermore, Murray-Clay et al. (2009) considered the evaporation for a hot jupiter with mass 0.7~M$_J$ and radius 1.4~R$_J$ with a pure UV radiation field at various fluxes. At high fluxes $\gtrsim 10^4$~\flux, they found the flow to be in approximate radiation-recombination equilibrium, in this case the flow behaved like a HII region, was fully ionized and approximately isothermal with a temperature of $10^4$~K and the cooling was dominated by Lyman-$\alpha$ radiation. As the density in any HII region scales with as $J_0^{1/2}$ (where  $J_0$ is the ionizing flux in photons per unit time), then the mass-loss rate in this regime scale approximately as $J_0^{1/2}$. At fluxes $\lesssim 10^4$~\flux the density in the flow was so low that recombination were too slow compared to the flow time and the mass loss rate then scaled linearly with flux. In this case ionizations were balanced approximately by advection, which Murray-Clay et al. 2009 identified as an ``energy-limited'' flow.

The physical understanding of the transition for ``recombination-limited'' to an ``energy-limited'' flow being that once the density drops in the flow such that the recombination time is long compared to the flow time then it can no-longer be in radiative-recombination equilibrium (e.g. Soker, 1999). This obviously depends on planet properties, through the flow time (e.g. planet radius and mass); and as such the transition flux below which the flow cannot be in radiative-recombination equilibrium is not going to be a  unique value. However, several studies of exoplanet evolution (e.g. Jin et al. 2014) take the results of Murray-Clay et al. (2009) for their single planet study and apply it to all planets which is likely to lead to widely inaccurate adopted mass-loss rates. 

In this paper we aim to study the properties of UV driven evaporation across the range of observed exoplanets, particularity at low masses which has as yet been unexplored. We aim to put forward a simple physical understanding of the various evaporation regimes and identify three: ``recombination-limited'', ``energy-limited'' and a new regime ``photon-limited''. We present an analytic picture in Section~2, perform a numerical study with a new radiation-hydrodynamic scheme in Section~3. In Section~4 we discuss our results and summarise in Section~5.

\section{Analytic Overview}
The evaporative flow problem is complicated by the fact that in general the flow may not be in local thermodynamic or ionization equilibrium, particularly at low fluxes (e.g. Murray-Clay et al. 2009). However, in the case of irradiation by a spectrum dominated by Hydrogen ionizing photons (EUV) we can make progress by considering ionization balance (e.g. Bertoldi \& Mckee, 1990). We consider a planet with mass $M_p$ and radius $R_p$ close to its central star at a separation $a$. Noting that $R_p/a\ll1$, then the EUV flux is constant on all scales of interest. We consider a flow that is dived into three distinct regions: (i) bolometrically heated planet's atmosphere; where the temperature is set by the flux of optical photons and the optical depth to EUV photons is large such that the ionization fraction ($X$) is very low. (ii) an ``ionization front'' where the photo-ionization rate greatly exceeds the recombination rate and the ionization fraction increases as one moves away from the planet. (iii) a fully ionized region where photo-ionization is balanced by recombinations and the ionization fraction is unity\footnote{Here we consider the case of a pure Hydrogen atmosphere, with the inclusion of other elements the ionization fraction is generally $>1$}. 

Since the scale height of the planet's bolometrically heated atmosphere must be $\ll R$ for a bound planet then the transition from the bolometrically heated atmosphere (region i) to the ionization front (region ii) is by construction sharp. In the picture we develop below we take it to occur at an optical depth $\tau_{\rm EUV}=1$ to the EUV photons and call this radius the planet's radius. However, unlike many problems in astrophysics involving EUV radiation, as we shall see we cannot in general assume that the ionization front is thin compared to the flow scale (or even the planet's scale) and in many cases we find a transition from region ii to region iii occurs at a very large radius. In particular, if it occurs at a radius larger than the sonic radius then it is irrelevant in determining the mass-loss and we assume the flow to only contain regions i and ii.    

We choose to work in one-dimensional spherically symmetric picture along a streamline connecting the star and planet (see Stone \& Proga, 2009 and Owen \& Adams, 2014 for the role of multi-dimensional flows). For the sake of clarity we neglect the contribution from the tidal gravity due to the star. For a steady-state, spherically symmetric flow the governing equations of hydrodynamics are:
\begin{eqnarray}
\frac{\partial}{\partial r}\left(\mu r^2nu\right)&=&0\label{eqn:mass}\\
u\frac{\partial u}{\partial r}{\bc +}\frac{1}{\rho}\frac{\partial P}{\partial r}{\bc +}\frac{GM_p}{r{\bc ^2}}&=&0\label{eqn:mom}\\
{\bc\frac{1}{r^2}}\frac{\partial}{\partial r}\left[{\bc u}r^2\left(\rho E+P-\frac{GM_p\rho}{r}\right)\right]&=&q\label{eqn:energy}
\end{eqnarray}
where $\mu$ is the mean-molecular weight given by $\mu=m_H/(1+X)$, with $X$ the ionization fraction; $G$ is the gravitational constant; $n$ is the number density {\bc of Hydrogen nuclei}; $\rho=n\mu$ the mass density; $u$ the gas velocity; $P=k_bnT$ the gas pressure, where $T$ is the temperature; $E$ is the {\it total} specific energy density and $q$ is the net volumetric heating rate, {\bc which is detailed in Section 3.1}. Furthermore steady state ionization balance requires {\bc (e.g. Bertoldi \& McKee 1990, their Equation 2.1)}:
\begin{equation}
J_0={\bc {\mathcal{F}_{\rm IF}}}-\int_\infty^{R_p}\!\!\!\!\!{\rm d}R \,\alpha_RX^2n^2\label{eqn:IB}
\end{equation}
where $\alpha_R$ is the recombination rate {\bc and $\mathcal{F}_{\rm IF}$ is the flux of ionizing photons into the ionization front. Since recombination are negligible inside the ionization front then $\mathcal{F}_{\rm IF}$ must balance the flux of neutral particles into the ionization front (Churchwell et al. 1987, Bertoldi \& McKee 1990). Namely:
\begin{equation}
\mathcal{F}_{\rm IF}=n(R_p)u(R_p) \label{eqn:PL1}
\end{equation}  
Note since the radiation field is plane parallel, Equation~\ref{eqn:IB} is written in plane parallel form. Since the flow is approximately spherical we must be careful combining Equation~\ref{eqn:IB} with the hydrodynamic terms, something we discuss further in Section 2.1.1. Specifically, we cannot trivially integrate Equation~\ref{eqn:PL1} over the surface of the planet by multiplying by $4\pi R_p^2$ to find the mass-loss rate, since the total number of ionizing photons entering the ionizing front does not necessarily equal the total number reaching the base of the atmosphere, as some may pass through the flow without interacting with the it.} Equation~\ref{eqn:energy} can be recast into a limiting mass-loss, the ``energy-limited'' mass-loss rate where uncertainties in heating and cooling are encapsulated into an efficiency parameter $\eta$ which can only be determined through numerical calculations with values typically in the range 0.01-0.3 (e.g. Murray-Clay et al. 2009; Owen \& Jackson, 2012).
\begin{equation}
\dot{M}\le\eta\frac{\pi R_p^3J_0h\bar{\nu}_h}{4GM_p}\label{eqn:EL}
\end{equation}
where $h\bar{\nu}_h$ is the mean photon energy that goes into heating (rather than ionization). {\bc The origin of the factor 4 in the denominator comes from averaging the received flux by a $\pi R_p^2$ disc over the $4\pi R_p^2$ surface area of the planet. We choose to this form as it is the commonly used form in the exoplanet literature and allows comparisons of the efficiency ($\eta$) values, although we note some authors do not perform such an averaging (e.g. Murray-Clay et al. 2009).} For a flow that can be considered close to being energy-limited requires that $q$ is predominately controlled by heating and radiative cooling is a small contribution. Therefore, energy losses are dominated by $P{\rm d}V$ work or advection. The main contribution to the radiative cooling rate $\Lambda$ is Ly$\alpha$ radiation with cooling rate (Black 1981):
\begin{equation}
\Lambda=7.5\times10^{-19}{\rm ~erg~cm^{-3}~s^{-1}}n_en_0\exp\left(\frac{{\bc -}118348~{\rm K}}{T}\right)
\end{equation} 
The ionization state of the gas is controlled by the recombination rate; adopting the on-the-spot approximation such that $\alpha_R=\alpha_B$ where $\alpha_B$ is the case-B recombination coefficient given by $\alpha_B=2.6\times10^{-13} {\rm ~cm^3~s^{-1}}(T/10^4~{\rm K})^{-0.7}$ so that the recombination rate $\Lambda_{\rm ion}$ is:
\begin{equation}
\Lambda_{\rm ion}=\alpha_Bn_e^2
\end{equation}
where $n_e$ is the electron density. Noting the ratio of the cooling time-scale ($t_c$) to the recombination time-scale ($t_{\bc r}$) is given by:
\begin{equation}
\frac{t_c}{t_r}=\frac{k_bT\Lambda_{\rm ion}}{\Lambda}
\end{equation} 
and is purely a function of temperature and ionization fraction. When the gas is close to $\sim10^4$~K and roughly in ionization-recombination equilibrium then $t_c/t_r\approx0.05$. However, if the flow were no longer in  {\it local} ionization equilibrium and the temperature falls below $10^4$~K. {\bc The temperature falls because the heating rate and ionization rate scale identically with the ionizing flux; therefore, since the ionizing photons are unable to obtain ionization equilibrium they are unable to heat the gas to $10^{4}~K$ temperature (i.e. the ratio of the heating to ionization time-scale is a constant). Thus, $P{\rm d}V$ cooling becomes the dominant cooling mechanism (e.g. Murray-Clay et al. 2009)}, so we find  $t_c/t_r\sim\exp(-10^5~{\rm K}/T)$. Thus, $t_c$ quickly becomes longer than the recombination time when the flow is no longer in {\it local} ionization equilibrium. Since {\it local} ionization {\bc balance} breaks down when when the recombination time becomes longer than the flow time-scale $t_{\rm flow}$, then one would expect that in the majority of the cases $t_c>t_{\rm flow}$ when $T<10^{4}$~K and the flow will be close to ``energy-limited'' (c.f. Soker 1999, Bear \& Soker 2011). This does not mean that Equation~\ref{eqn:EL} will provide an accurate description of the mass-loss rate over the range of parameters of interest as $\eta$ may still vary considerably due to hydrodynamic effects.  

\subsection{Limiting cases}

Inspection of Equation~\ref{eqn:IB} \& \ref{eqn:EL} indicate that three physical {\bc processes} may limit the mass-loss. Either the number of incoming photons, recombinations or energy considerations. Equating $J_0$ with the advection term in Equation~\ref{eqn:IB} indicates that the flow will be limited by the number of incoming ionization photons --- a processes we term ``photon-limited". Equating $J_0$ with the recombination term in Equation~\ref{eqn:IB} results in a density in the ionization front that is determined only by radiation-recombination balance --- a processes we will term ``recombination-limited". Finally, insisting that the flow satisfies Equation~\ref{eqn:EL} will result in a flow that is limited by the available energy, this has become known commonly as ``energy-limited'' evaporation. In this case the mass-loss rate is described by taking Equation~\ref{eqn:EL} as an equality where $\eta$ is yet to be determined.

\subsubsection{Photon limited flows}\label{sec:photon_lim}

This corresponds to the case where the ionization front encompasses the entire flow (including the sonic point) and as such recombinations are completely negligible in the flow {\bc  so that $\mathcal{F}_{\rm IF}=J_0=n(R_p)u(R_p)$}. {\bc Additionally the pull of gravity is weak and the inequality in Equation~6 is easily satisfied. Note we are implicitly assuming that the gas temperature is high enough that the gas is able to escape freely.} . Therefore, we are limited by the number of incoming ionizing photons: namely we can only loose one atom per incoming photon. {\bc Note we are implicitly assuming that the gas temperature reached in the ionization front is high enough that escape from the planet's potential is possible; if not the flow becomes ``energy-limited'' as we will discuss in Section~2}. However, here we are complicated by geometric effects. For uniform spherical irradiation then the total number of absorbed photons is the area  out to the ionization front, not the planet's radius itself. However, planet evaporation does not occur with spherical irradiation, but rather plane parallel. Furthermore, since the ionization front is marginally optically thin ionizing photons, then any photons with an impact parameter much larger than the planets radius, but smaller than the ionization front radius are likely to pass straight through the flow. Therefore, we can approximate the effective absorbing area of the planet as $\sim \pi R_p^2$. Therefore, in order to obtain the photon limited mass-loss rate we drop the recombination integral in Equation~\ref{eqn:IB} and equating $J_0$ and the advection allows us to calculate the mass-loss rate approximately as:
\begin{equation}
\dot{M}_{\rm PL}\approx \pi R_p^2m_HJ_0\label{eqn:mdot_PL}
\end{equation}  
We note that we do not need to make assumptions about the energy considerations here as we do not need to know the gas temperature to find the mass-loss rates. Such a mass-loss limit has not previously been discussed in the context of planet evaporation. Thus,  it will corresponds to cases where planetary gravity is not strong and the flux is low, so namely low-mass planets at large separations and early times. Unfortunately, due to the fact any 1D model essentially assumes spherically symmetric irradiation any photon limited mass-loss rate is likely to be much higher than this value (since the effective absorbing area in such a calculation is $\pi R_{\rm IF}^2$ and $R_{\rm IF}> R_p$ often with $R_{\rm IF}\gg R_p$). Therefore, in our numerical calculations presented in Section~3 we will not be able to study this limit of planet evaporation. Any numerical constraints on ``photon-limited'' evaporation must be obtained with 2/3D numerical calculations. Although, Equation~\ref{eqn:mdot_PL} should roughly represent an appropriate estimate for the mass-loss rate, {\bc  and is similar to the 1+1D numerical calculations of Bertoldi \& McKee (1990)}. 

\subsubsection{Recombination limited flows}
``Recombination-limited'' flows are perhaps the easiest to understand as in this case the radiative transfer and hydrodynamics problems are essentially de-coupled and the basics in the context of planetary evaporation were discussed in Murray-Clay et al. (2009). When the recombination time is short compared to the flow time-scale the ionization front becomes thin compared to the flow scale and essentially the transition from region i to region iii can be considered infinitesimally thin. Furthermore, the gas is thermostated to approximately $10^4$~K  and approximately isothermal. For cases when the planet's gravity is strong enough that the scale height in $10^4$~K is small compared to the planet's radius then one can approximate the density structure close to the planet as hydrostatic and the integral in Equation~\ref{eqn:IB} becomes (see also Murray-Clay et al. 2009):
\begin{equation}
\int_\infty^{R_p}\!\!\!\!\!{\rm d}R \,\alpha_RX^2n^2\approx\alpha_Bn_b^2\left(\frac{c_s^2R_p^2}{2GM_p}\right)\label{eqn:sg}
\end{equation}
where $n_b$ is the density at the ionization front. However, in the case that the planet's gravity is weak then the density profile will be roughly that of a constant velocity outflow and will fall of as $r^{-2}$. Therefore, in this case the integral in Equation~\ref{eqn:IB} becomes (see Johnstone et al. 1998):
\begin{equation}
\int_\infty^{R_p}\!\!\!\!\!{\rm d}R \,\alpha_RX^2n^2\approx\alpha_Bn_b^2\left(\frac{R_p}{3}\right)\label{eqn:wg}
\end{equation}
Writing the term in parenthesis in Equations~\ref{eqn:sg} and \ref{eqn:wg} as a scale height $H$ then we crudely set:
\begin{equation}
H=\min\left[\frac{R_p}{3},
\left(\frac{c_s^2R_p^2}{2GM_p}\right)\right]
\end{equation}
Therefore,  we can write the base density as:
\begin{equation}
n_b=\sqrt{\frac{J_0}{\alpha_BH}}
\end{equation}
Now we can write the recombination-limited mass-loss rate, noting in fully ionized flow $\mu=m_H/2$ as:
\begin{equation}
\dot{M}_{\rm RL}=2\pi R_p^2m_Hn_bc_s\mathcal{M}_b\label{eqn:mdot_RL}
\end{equation}
where $\mathcal{M}_b$ is the Mach velocity at the base of the flow. For an isothermal flow $\mathcal{M}_b$ is given by the well known Parker wind solution (Parker, 1958, Cranmer 2004) where:
\begin{equation}
\mathcal{M}_b=\sqrt{-W_0\left[-\frac{R_p}{R_s}^{-4}\exp\left(3-\frac{4R_s}{R_p}\right)\right]}\label{eqn:mach}
\end{equation} 
with $W(x)$ the Lambert $W$ function and $R_s=GM_p/2c_s^2$, if $R_p/R_s<1$ and $\mathcal{M}_b=1$ otherwise. For tightly bound planets Equation~\ref{eqn:mach} has an exponential fall off of the form:
\begin{equation}
\mathcal{M}_b\approx\left(\frac{R_p}{R_s}\right)^{-2}\exp\left(-\frac{2R_s}{R_p}\right)
\end{equation}
Therefore for $R_p/R_s$ close to unity $\dot{M}_{RL}\propto R^{3/2}$; however for $R_p/R_s\ll 1$ it has a steep (exponential) dependence on mass and radius.

\subsection{Picture of Evaporation across parameter space}

For fixed mass, since $\dot{M}_{\rm EL}\propto\eta R{\bc _p}^3$ and $\dot{M}_{\rm PL}\propto R_{\bc p}^2$, ({\bc i.e. they intercept once}) then we note that there are two possible cases. Either the flow is ``recombination-limited'' at all possible radii (high masses and high fluxes) or the flow is ``recombination-limited'' at both large radii (where $\dot{M}_{RL}\propto R_{\bc p}^{3/2}$) and at small radii where it has an steep dependence on radius. Furthermore, since ``photon-limited'' mass-loss has a shallower dependence on radius than ``energy-limited'' then photon limited flows will dominate at larger planetary radii and lower masses than ``energy-limited''. The transition between ``photon-limited''  and ``energy-limited'' flows can be found by taking Equation~\ref{eqn:EL} as an equality and equating it with Equation~\ref{eqn:mdot_PL} to find:
\begin{equation}
M_p=\eta\left(\frac{h\bar{\nu}_h}{4Gm_h}\right)R_p
\end{equation} 
{\bc Note this expression indicates that when the photon energy is much larger that the depth of the gravitational potential, then the photon limit is the relevant one at low fluxes (i.e. we can assume heating is sufficient to allow the atmosphere to escape as we adopt in Section 2.1.1). Whereas, when the photon energy is much less than the depth of gravitational potential, the energy limit is the relevant one at low fluxes as one cannot assume the gas can escape freely.}

One can find the transition between ``recombination-limited'' and both ``photon-limited'' and ``energy-limited'' by equating Equation~\ref{eqn:mdot_RL} with Equations~\ref{eqn:mdot_PL} \& \ref{eqn:EL} respectively to find:
\begin{equation}
W_0\left[-\frac{R_p}{R_s}^{-4}\exp\left(3-\frac{4R_s}{R_p}\right)\right] = - \frac{J_0\alpha_B H}{4c_s^2}\label{eqn:tran1}
\end{equation}
for the transition between ``recombination-limited'' and ``photon-limited'' and
\begin{eqnarray}
W_0\left[-\frac{R_p}{R_s}^{-4}\exp\left(3-\frac{4R_s}{R_p}\right)\right] =&& -\eta \frac{J_0\alpha_B H}{4c_s^2}\nonumber \\&\times&\left(\frac{R_ph\bar{\nu}_h}{GM_pm_h}\right) \label{eqn:tran2}
\end{eqnarray}
for the the transition between ``recombination-limited'' and ``energy-limited'' where we note the last term in brackets is the ratio of the photon energy to binding energy. Equations \ref{eqn:tran1} \& \ref{eqn:tran2} posses either two solution or no solutions and cannot be presented in closed form. Therefore, we proceed graphically.  

\begin{figure*}
\centering
\begin{tabular}{cc}
\includegraphics[width=0.5\textwidth]{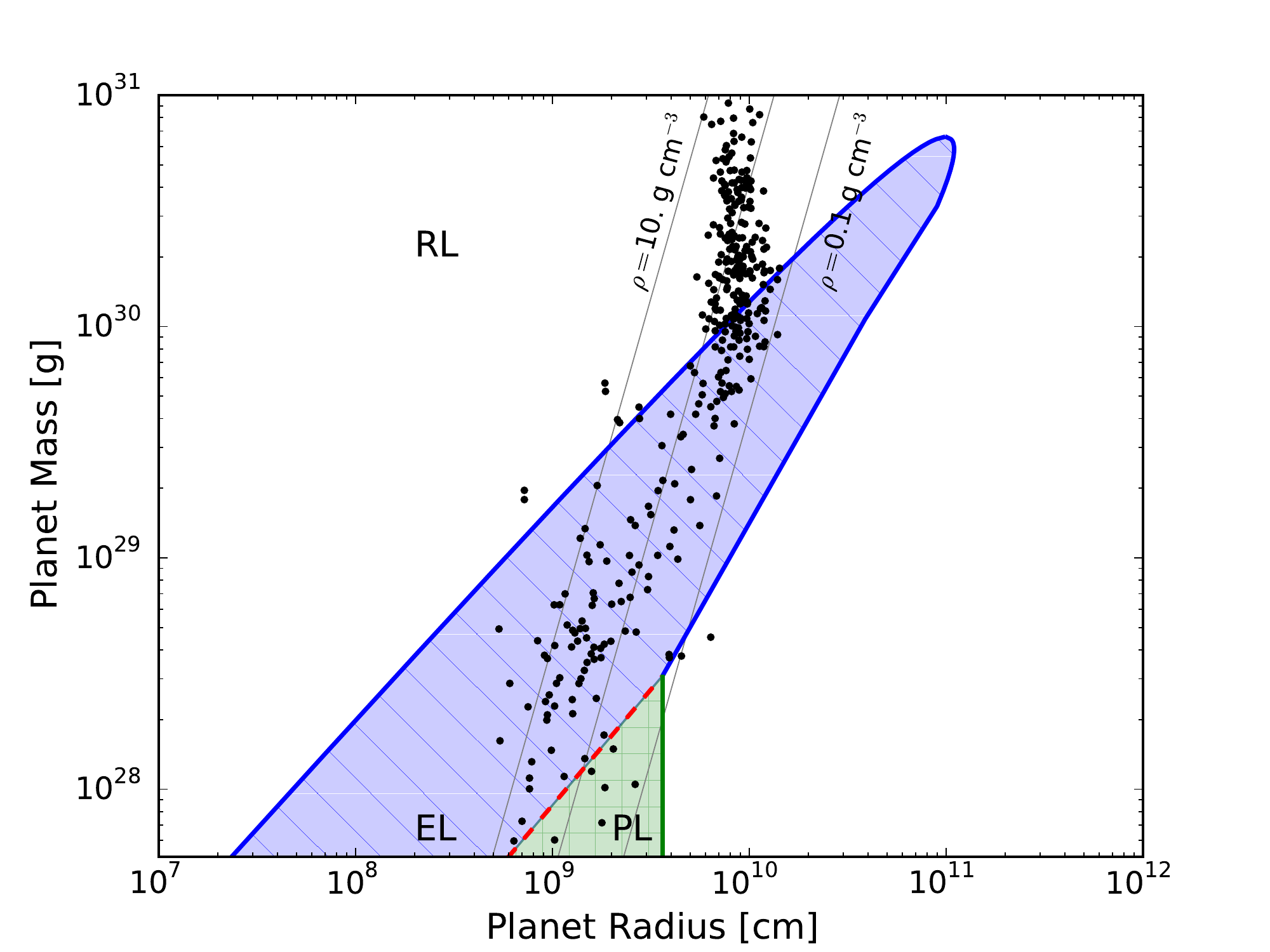} &
\includegraphics[width=0.5\textwidth]{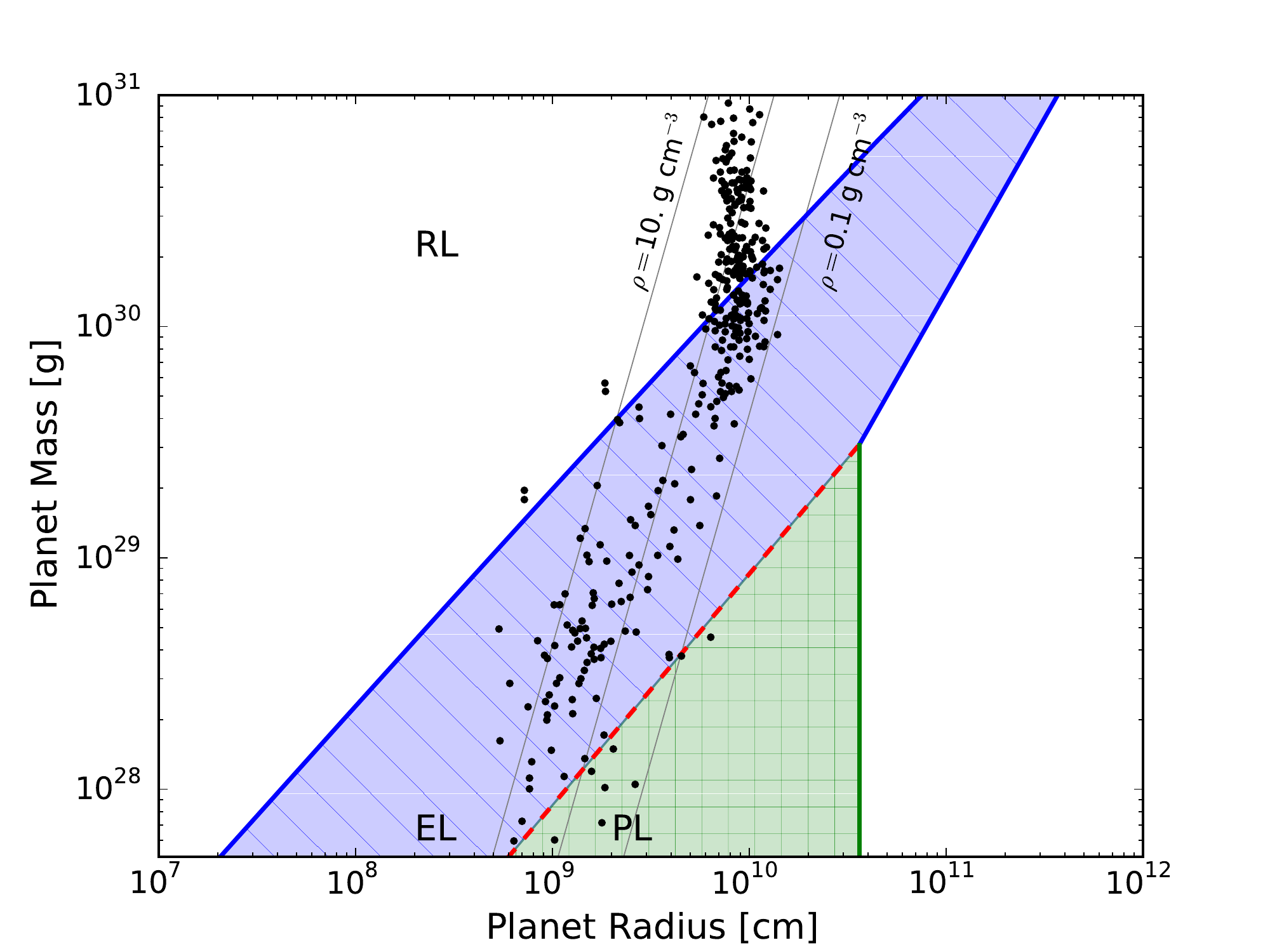} \\
\includegraphics[width=0.5\textwidth]{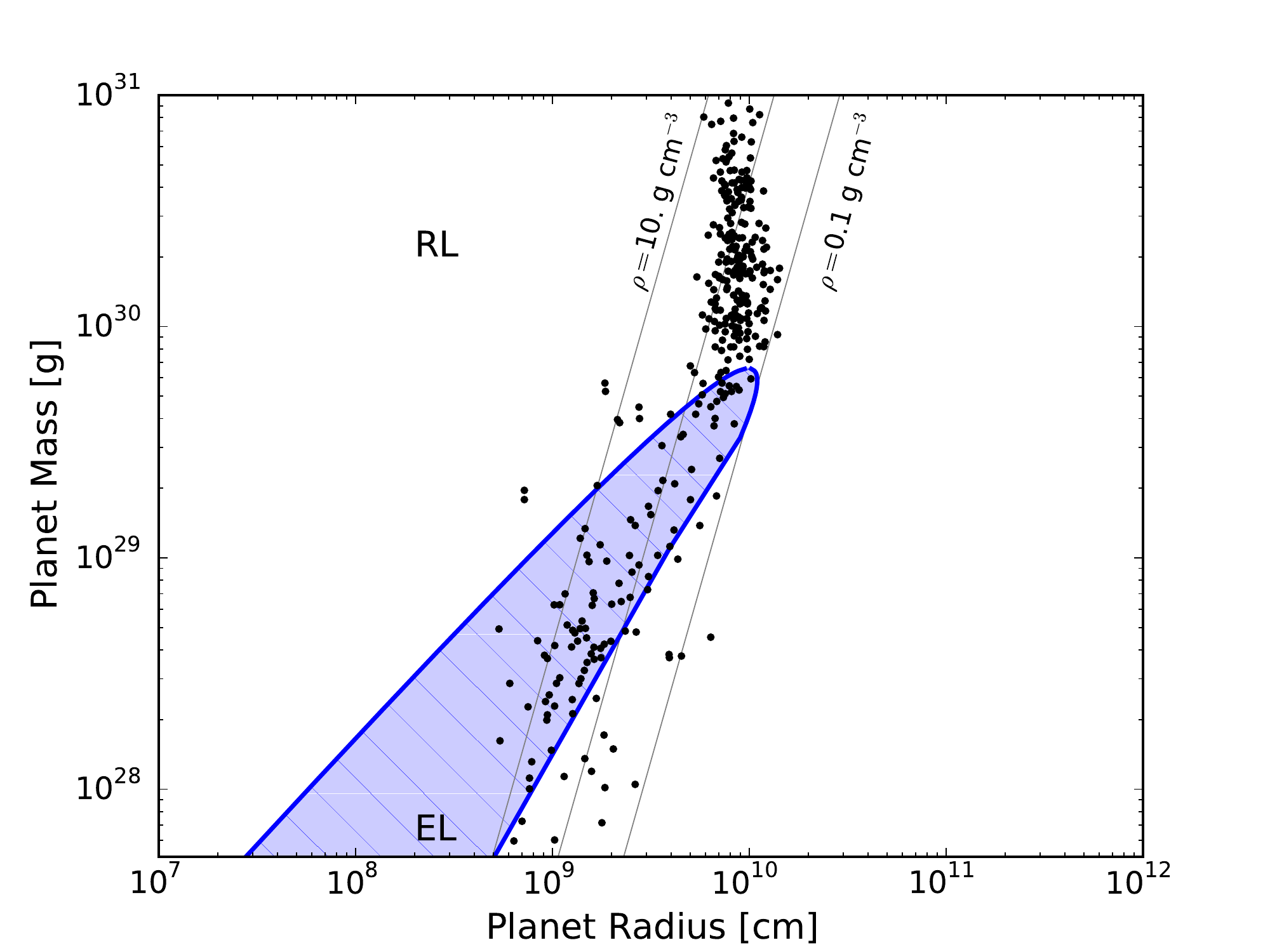} &
\includegraphics[width=0.5\textwidth]{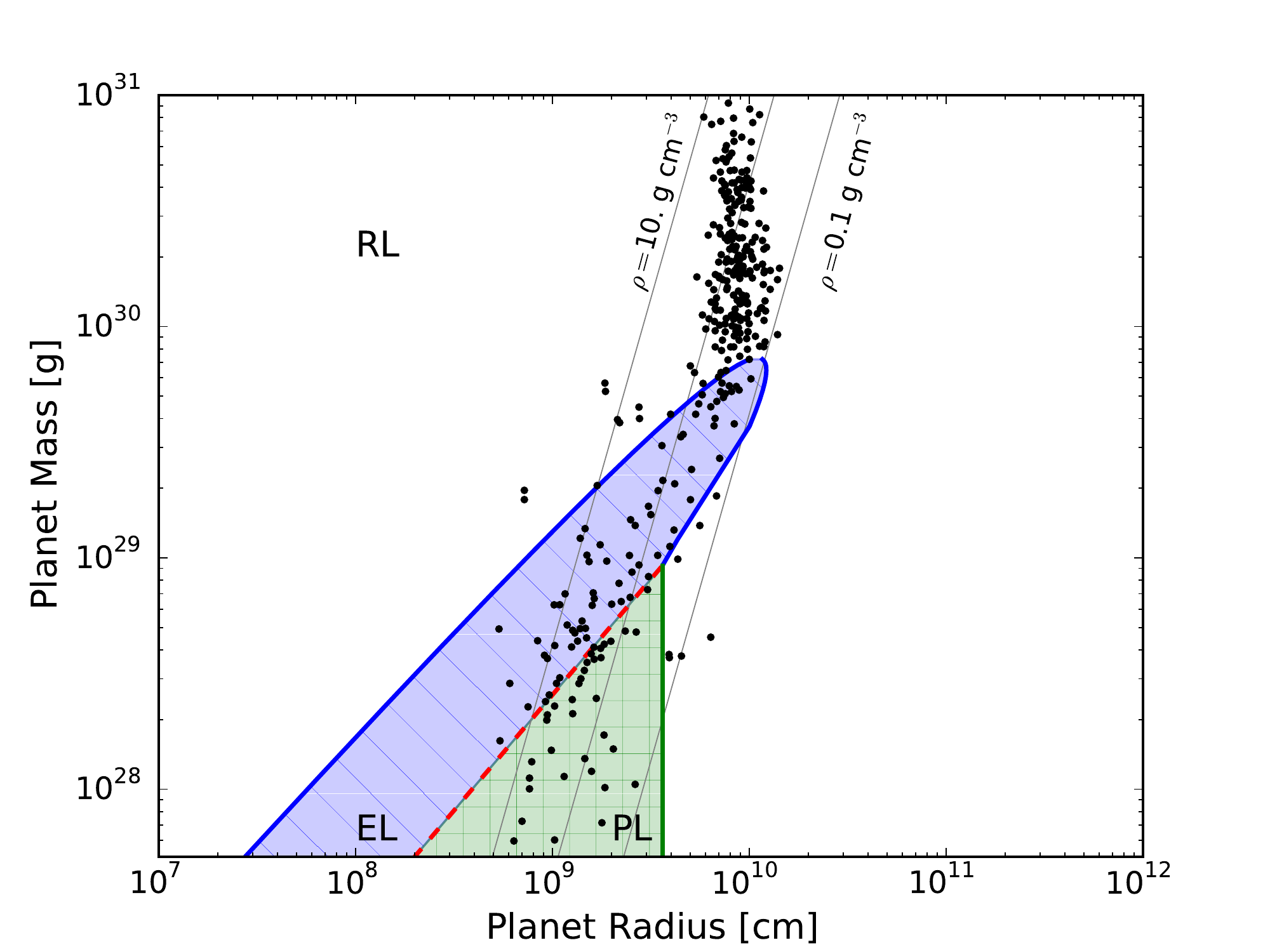}
\end{tabular}
\caption{The various mass-loss regions are shown as a function of planet radius and mass for various incident EUV fluxes and mass-loss efficiencies. The crossed green region shows ``photon-limited'' mass-loss, the blue striped region show ``energy-limited'' mass-loss and the remaing parameter space is recombination limited. The thin lines show lines of constant density at 0.1, 1.0 \& 10 g cm$^{-3}$ and the points show currently observed exoplanets obtained from the open exoplanet catalogue - data obtained on 5th February 2015 - (Rein, 2012). The top left panel shows our fiducial case for a flux of $3\times10^{4}$~\flux and efficiency $\eta=0.1$. The bottom left and top right panels show fluxes of $3\times10^{5}$ \&  $3\times10^{3}$~\flux respectively. The bottom right panel shows a flux of $3\times10^{4}$~\flux and efficiency $\eta=0.3$. }\label{fig:regions}
\end{figure*}

In computing the various mass-loss regimes we assume that the mass-loss efficiency is constant with planet radius and mass. As lowering the efficiency reduces the energy limited mass-loss rate this makes this limit more restrictive and increases the range of parameters over which the flow can be characterised by energy-limited evaporation. For our fiducial case we take $h\bar{\nu}_h=6.4$~eV, $\eta=0.1$ and use an EUV flux of $3\times10^{4}$~\flux and the various mass-loss regimes are shown in top left panel of  Figure~\ref{fig:regions}. In our panels of Figure~\ref{fig:regions} we also show a recent census of the exoplanet population as points. 

This moderate flux corresponds roughly to a separation of 0.2~AU around a young star and 0.04~AU for a 1~Gyr old star (Jackson et al. 2012). In this example we see that jupiter mass and above planets are evaporating in the recombination limit, whereas most sub-jupiter mass planets are losing mass in an energy limited fashion. For very low mass planets ($\sim$M$_\oplus$) we find that evaporation is photon limited. We note that while most planets evolve to the left in the diagram as they shrink this is only a minor adjustment for massive planets. However, planets with envelope mass-fractions of order 10s of percent can be much larger at early times, and can often have radii 5-10 times large than at Gyr ages when they are observed today (Lopez \& Fortney 2013; Owen \& Wu 2013).

In the bottom left and top right panels of Figure~\ref{fig:regions} we explore the role incident flux plays, with fluxes of $3\times10^{3}$ and $3\times10^5$~\flux respectively. As one expects at lower fluxes the parameter space dominated by energy-limited and photon-limited flows, with the majority of planets loosing mass in an energy limited fashion and a small minority in a photon limited picture. While massive ($>{\rm M}_J$) planets still remain in the recombination limit. 

Alternatively at high fluxes no planets are in the photon-limited picture and the energy limited flow occupies a small fraction of parameter space and most planets are evaporating in the recombination limited scenario. Since at this flux the majority of planets that sit in the energy limited-region are low mass. In order to experience fluxes of this magnitude requires the planets to been irradiated at early time ($<100$~Myr - Jackson et al. 2012, Owen \& Wu 2013) and as such the planets plotted here would have been considerably larger and most would sit to the right of the energy-limited region.

Finally, we investigate how the mass-loss efficiency parameter in the energy-limited flows affects the parameter space. In the bottom right panel of Figure~\ref{fig:regions} we show our fiducial model, but with an efficiency parameter of $\eta=0.3$ corresponding to the upper range found by Owen \& Jackson (2012). As expected increasing the efficiency parameter reduces the importance of the energy-limit and the majority of low mass planets are now in the photon-limit. Whereas energy-limited evaporation is constrained to intermediate mass-planets.

The picture described here is consistent with the study of Murray-Clay et al. (2009) who studied a 0.7~M$_J$ planet with a radius of $10^{10}$~cm as a range of fluxes and found that the flow transitions from recombination limited at high fluxes to energy limited at low fluxes around $\sim10^{4}$~\flux, consistent with our calculations that predict a transition flux of $\sim 2\times10^{4}$~\flux. 

Now we can build up a picture for the mass-loss histories of both low mass planets ($\lesssim 20$M$_\oplus$) and high mass planets ($\sim$M$_J$). Close-in high mass planets begin evaporating in the recombination limited regime, as they evolve and the EUV flux falls with time they may transition to energy limited at some low flux, for typical parameters this is about $10^4$~\flux; however, we note that this is a strong function of planet mass and radius (as the transition occurs when $R_p/R_s\ll1$). Similarly, close-in low mass planets will begin evaporating in the recombination limit as they contract significantly due to cooling and the flux drops they may make a transition to photon-limited evaporation; however as the flux drops further and they contract even more they will eventually transition to energy limited evaporation. Whether or not low-mass planets experience photon-limited evaporation and when they make the transition from recombination to photon to energy limited is non-trivial. What is certain is there is not a unique value of the flux that one transitions from one regime to the other as assumed in previous evolutionary studies (e.g. Jin et al. 2014). Furthermore, our analysis presented here makes no effort to quantify what the efficiency of energy limited flow is and that can only be answered by numerical calculations such as those we preform in the next Section.

\section{Numerical Simulations}

In the previous section we have put forward the basic physical picture of planetary evaporation driven by EUV radiation, and their scalings. However, the actual mass-loss rates are still unknown as the details of the picture remain unquantified. In particular, the efficiency of any energy limited flow, or the exact temperature profile of any radiation recombination limited flow is sensitive to the exact details of the heating and cooling.

Furthermore, since we only expect the ionization front to be thin in the case of a flow which is recombination limited, and as we shall see it often the scale of the planet, one cannot assume a monochromatic irradiation field (as done by Murray-Clay et al. 2009) and radiation hardening will need to be taken into account. Thus, the only way to obtain an accurate picture of planetary evaporation we must proceed and perform numerical calculations. 

We follow many other previous approaches and consider 1D spherically symmetric outflow, where we solve the problem along a streamline connecting the star and planet, neglecting the Coriolis force. Along this streamline we solve the radiation-hydrodynamic problem including EUV irradiation, and non-equilibrium heating/cooling and Hydrogen ionization chemistry. We use the {\sc zeus-MP} hydrodynamics code (Stone \& Norman, 1992, Hayes et al. 2006), which has been previously used for planetary evaporation studies (Owen \& Adams, 2014) and add to it a new module to perform the radiative-transfer, ionization structure and heating and cooling update. 

As discussed in Section~\ref{sec:photon_lim}, our numerical setup will result in a ``photon-limited'' mass-loss rate that is too high compared to reality. In fact in all simulations we perform we find that the photon limited rate of our spherically symmetric setup is much higher than any measured mass-loss rates. As such in our numerical study here we only investigate the transition between energy and recombination limited and we must leave an investigation of the transition to photon limited flows to future multi-dimensional calculations. 

\subsection{Ionizing radiative transfer}
Ionizing radiative transfer can be solved most efficiently in a causal manner where one performs the ray tracing calculation proceeding away from the source (towards the planet in our case) in a step-wise manner. We operator split our radiative transfer calculation from the hydrodynamics and perform it between the ``source'' and ``transport'' steps of the standard {\sc zeus} scheme. We consider only pure Hydrogen and the equations for ionization and energy due to this update are given by:
\begin{eqnarray}
\frac{{\rm d}X}{{\rm d}t}=(1-X)(\Gamma_i+n_eC_H)-Xn_e\alpha_B \label{eqn:ion_frac} \\
\frac{\partial e}{\partial t}=-P\nabla\cdot{\bf u} +\Gamma_h-\Lambda \label{eqn:eng}
\end{eqnarray} 
where $\Gamma_i$ is the photoionization rate, $C_H$ is the collisional ionization rate, $e$ is the internal energy of the gas, $\Gamma_h$ is the heating rate and $\Lambda$ is the cooling rate. We include heating from photo-ionizations and cooling from recombination, collisional excitation, collisional ionization and free-free emission (Katz et al. 1996). {\bc However, we neglect heat transport due to conduction as it known to be sub-dominant by many orders of magnitude (e.g. Murray-Clay et al. 2009)}. We follow (Mellema et al. 2006) and pre-tabulate the heating and ionization rates as a function of neutral hydrogen column density for our input spectrum, thereby capturing the non-monochromatic nature of our problem. Equations~\ref{eqn:ion_frac} \& \ref{eqn:eng} are then solved explicitly by sub-cycling then on time-scales which can be shorter than the hydrodynamic time-step. We note that our scheme requires that we resolve the ionization front, something we have enough resolution to do. 

\subsection{Numerical setup}
We perform a series of simulations in order to better understand the regimes which planetary evaporation exists in. Since a full exploration of the entire parameter space numerically is beyond the scope of this work, we instead choose to focus on low-mass planets (where evaporation is more evolutionary important - see Owen \& Wu 2013; Lopez \& Fortney 2013). In all simulations we use an underlying bolometrically heated atmosphere of 1000~K, which is forced to be isothermal by means of a temperature floor. We adopt a grid with 192 non-uniformly spaced cells with a resolution of $\sim$20~km at the base of the grid, and a radial extent from $R_p$ to $30~R_p$. The simulation is initiated with a neutral hydrostatic density structure for a 1000~K atmosphere. {\bc At the lower boundary the ghost cells are reset to the neutral, hydrostatic density and temperature structure every time-step. At the upper boundary we use standard outflow boundary conditions, which are perfect in 1D super-sonic outflow (see Stone \& Norman 1992; Hayes et al. 2006, for their implementation in the {\sc zeus} code). In order to choose the density at the inner boundary we set the density, such that the initial neutral hydrostatic atmosphere has an optical depth of $\tau=100$ to 13.6eV photons, which typically results in densities of order $\sim 10^{-11}-10^{-12}$g~cm$^{-3}$. We find that such a choice is sufficient to keep the inner regions of the active domain part of the bolometrically heated atmosphere rather than UV heated atmosphere.}  The domain is then irradiated by the UV and a  transonic thermal wind is launched. In all cases the flow approaches steady-state and no variability is found. All results are analysed after $\sim 100$ sound crossing times after steady-state is firmly established, {\bc with the mass-flux constant at the $\sim 10^{-3}$ level (see Figure~2)}.

We perform two series of simulations: one at constant planet mass with a mass of 10~M$_\oplus$ where we vary the planet radius and one set at constant radius of 3.14~R$_\oplus$ where we vary the planet's mass. In both cases the radius/mass are varied such that the smallest/most massive planets have a density of 10~g~cm$^{-3}$ and the largest/least massive have at ratio of planet radius to the ``Parker'' radius\footnote{The ``Parker'' radius is the sonic radius of an isothermal flow, where we take the temperature of the underlying bolometrically heated atmosphere and is given by: $GM_p/2c_s^2$; planets larger than this radius are not hydrostatically bound over evolutionary time-scales.} of 0.2. For each set we perform 24 for individual simulations at {\bc 12 different fluxes varying between $\sim 10$ and $\sim10^6$~\flux (where we take the EUV luminosity contained between 13.6 and 30eV)}. Our input spectrum was calculated by Ercolano et al. (2009) and has been used in previous exoplanet evaporation studies by Owen \& Jackson (2012) and extends out to several 100eV (although photons with high energies have an incredibly small photo-ionization cross-section with Hydrogen, $\sigma\propto E^{-3}$).  

\subsection{Results}

\begin{figure*}
\centering
\begin{tabular}{c}
\includegraphics[width=1.\textwidth]{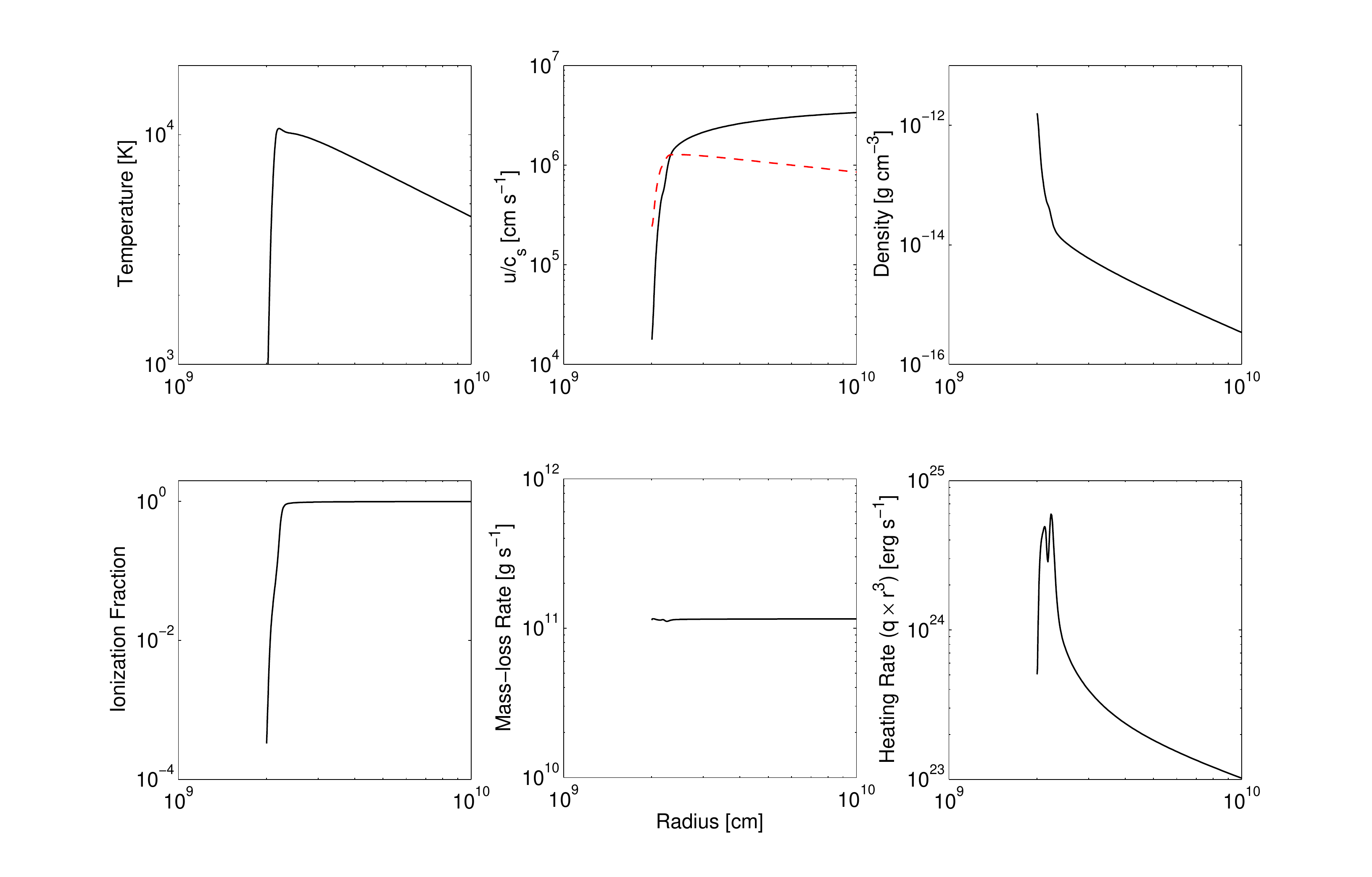}\\
(a)\\
\includegraphics[width=1.\textwidth]{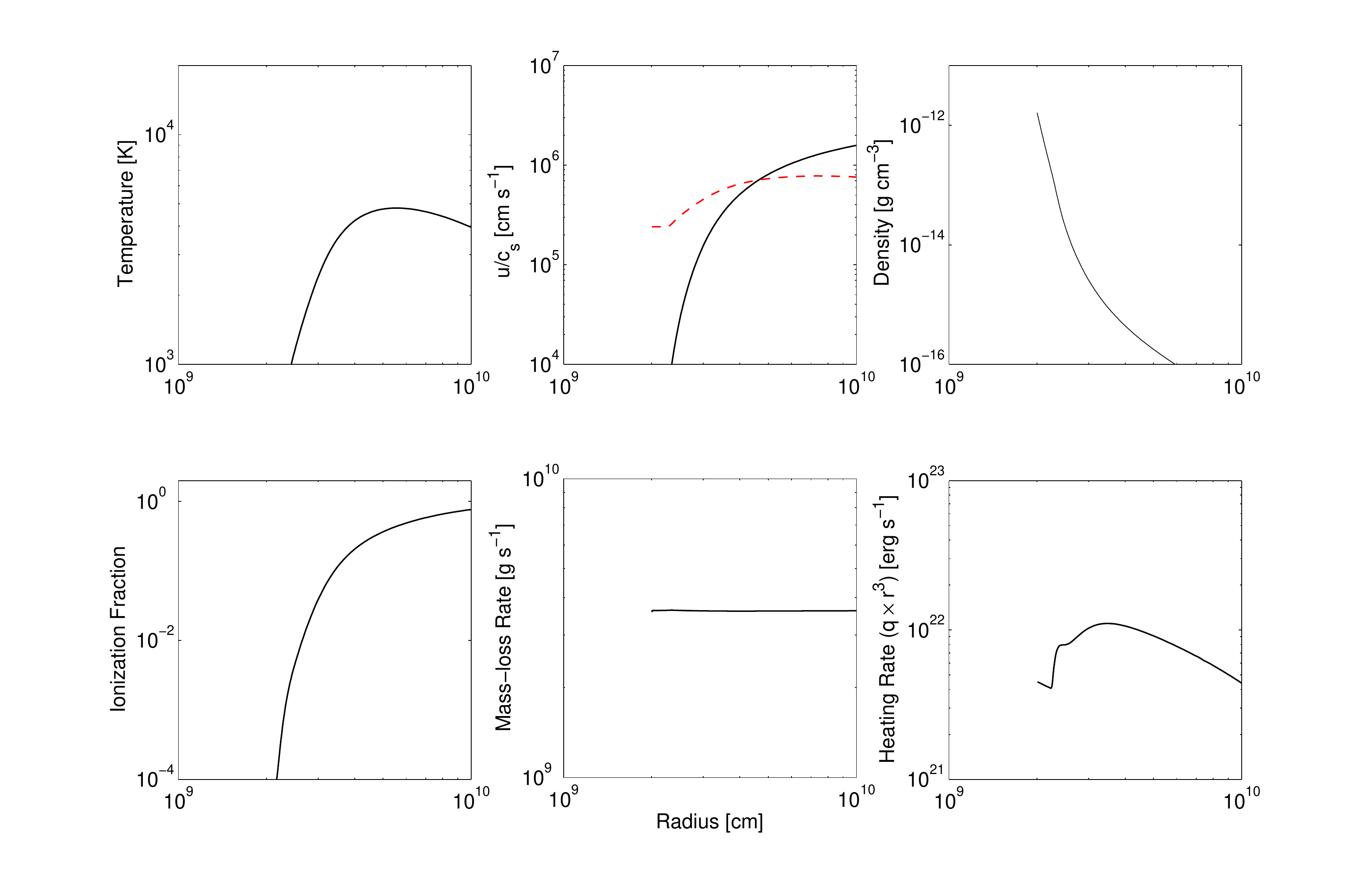}\\
(b)
\end{tabular}
\caption{\bc Flow toplogy for a planet with radius of 3.14~R$_\oplus$ and mass of 6.25~M$_\oplus$ receiving an ionizing flux of $10^{6}$~\flux (a) and $10^{3}$~\flux (b). The top left panels show the gas temperature; bottom left panels show the ionization fraction; top middle panels show the velocity (solid) and sound speed (dashed); bottom middle panels show the mass-flux (i.e. $4\pi r^2 \rho u$); top right panels show the density and bottom right panels show the volumetric heating rate ($q$) multiplied by $r^3$ to indicate where most of the heating is taking place.}\label{fig:topology}
\end{figure*}

As expected we identify two limiting flow types: recombination limited at high fluxes and energy limited at low fluxes. Note as discussed above ``photon-limited'' flow cannot be found for realistic parameters in 1D spherically symmetric simulations. We show the different flow topologies for recombination limited (top panels) and energy limited flow (bottom panels) in Figure~\ref{fig:topology}. Both simulations are for a planet with a radius of 3.14~R$_\oplus$ and mass of 6.25~M$_\oplus$. The recombination limited flow has a flux of $10^{6}$~\flux and the energy limited flow has a flux of $10^{3}$ ~\flux. 

In the ``recombination-limited'' case (top panels), we see that the size of the ionization front is small compared to the planet's radius and the flow is fully ionized at large radius indicating recombination-ionization equilibrium. The temperature quickly reaches $10^{4}$~K at the ionization front and the sonic point is very close to the planet's surface. {\bc Furthermore, we note that most of the heating is concentrated close to the planet's surface, as indicated by the heating rate profile ($qr^3$ - multiplied by $r^3$ to indicate the radius where the majority of heating takes place) shown in the bottom right panel of Figure~2a.} We find that at these low masses the flow does adiabatically cool slightly at large radius although this is in the super-sonic regime of the flow it doesn't affect the mass-loss rate when compared to the simple estimates presented above.

In the energy limited case (bottom panels), we find that the ionization front essentially encompasses the entire flow and the flow is still partly neutral at 10 planetary radii. This results in a much colder flow only reaching $\sim 5000$~K at 2 planetary radii, before adiabatically cooling at large radii. This results in the sonic point being pushed to a larger radius. {\bc This flow topology is emphasised by the fact the heating rate peaks at larger radii than in the recombination limited case and is more spatially extended, as shown in the bottom right panel of Figure~2b.}  

\begin{figure}
\centering
\includegraphics[width=\columnwidth]{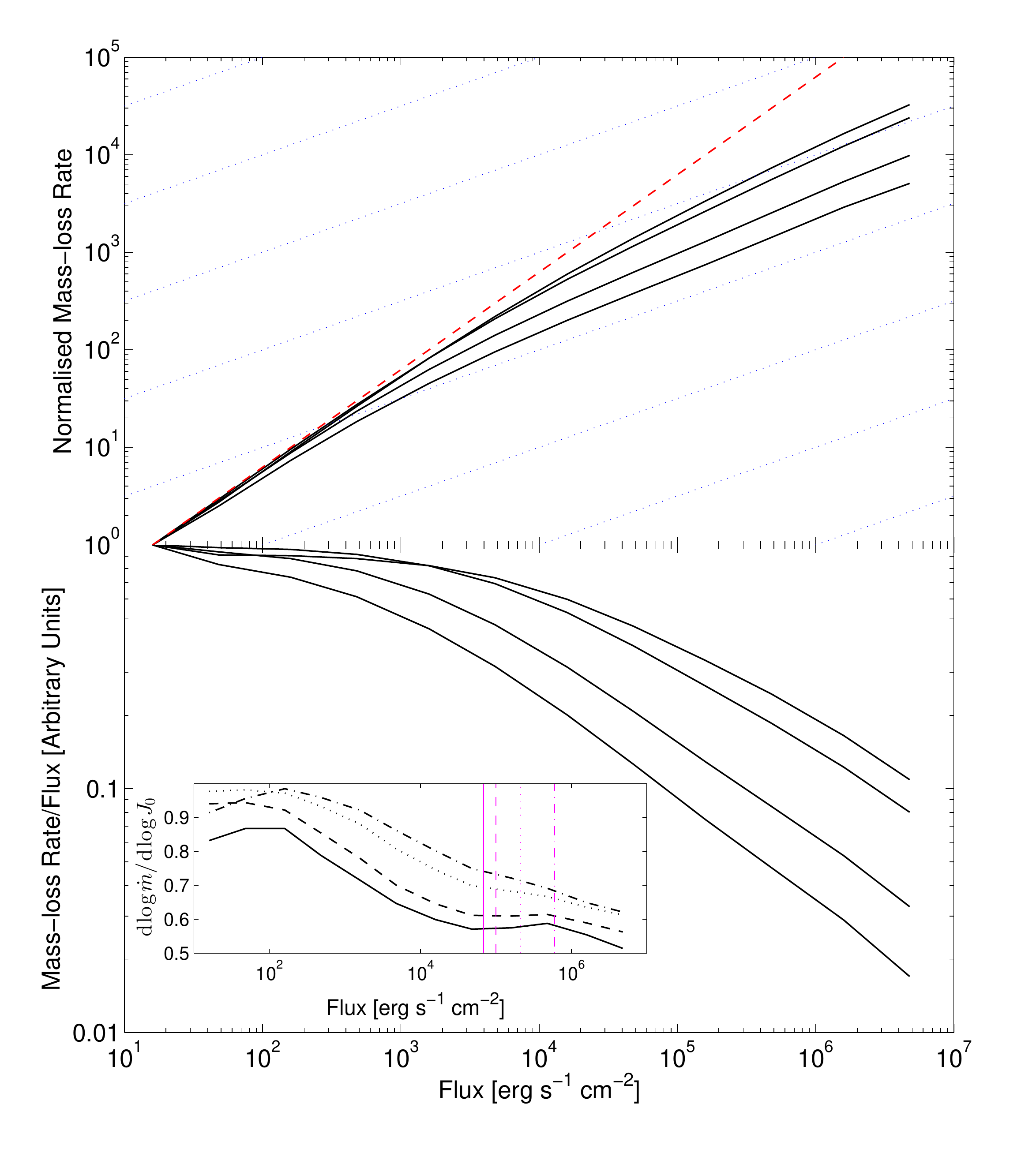}
\caption{The top panel shows the normalised mass-loss rate (the mass-loss rate for all cases is scaled to equal unity at the lowest flux of $\sim$10~\flux) as a function of flux shown as solid lines for a planet with a radius of 3.14~R$_\oplus$ and masses of (from top to bottom) 5.12, 6.25, 10.31, 13.92 M$_\oplus$. The dashed line shows the scaling expected for a pure ``energy'' limited flow. The dotted line show lines propotional to the square root of flux, the scaling expected from recombination limited flows. The bottom panel shows the same mass-loss rates divided by flux (again scaled so that the ratio of mass-loss rate and flux is one at the lowest flux of $\sim$10~\flux. The insert shows the logarthmic derivative of mass-loss rate by flux at the different masses 5.12 (solid), 6.25 (dashed), 10.31 (dotted), 13.92 (dashed) M$_{\oplus}$, the vertical lines show the analytical parscription would predict the flow transitions from energy limited to recombination limited.}\label{fig:tran}
\end{figure}

We can investigate the transition between energy-limited and recombination limited in Figure~\ref{fig:tran} where we plot the normalised mass-loss rate (where the mass-loss rate for all models is normalised such the mass-loss rate is unity at the lowest flux of $\sim 10$~\flux) against flux (top panel) and mass-loss rate divided by flux (bottom panel) for a planet with several different masses: 5.12, 6.25, 10.31, 13.92~M$_\oplus$ (with a radius of 3.14~R$_\oplus$). In the insert we show the logarithmic derivative of mass-loss rate with flux. At low fluxes the mass-loss rate scales linearly with flux following the dashed line as expected for energy limited flow; however, at large fluxes the mass-loss rate scales approximately as the square root of flux (following the dotted lines), indicating recombination limited flow. We find that the transition point varies with mass and perhaps more interestingly the transition is slow, taking a large range in flux.The predicted transition fluxes from Section~2 -- as indicated by the vertical lines in the insert of Figure~\ref{fig:tran} -- pick out the end of the transition. As expected planets with higher masses become recombination limited at lower fluxes due to there increased flow-time scales. 

\begin{figure}
\centering
\includegraphics[width=\columnwidth]{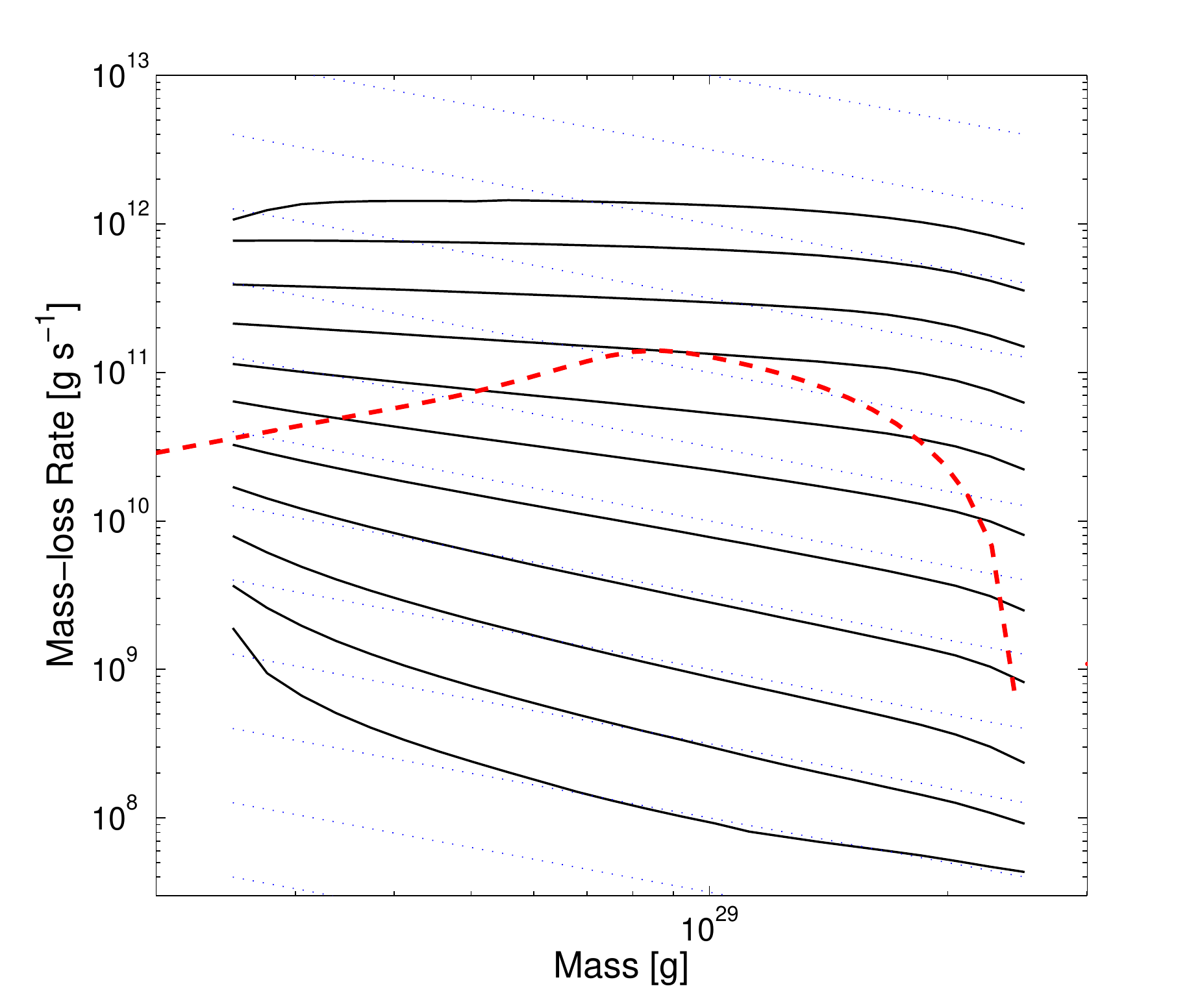}
\caption{Mass-loss rate as a function of planet mass for the full ranges of fluxes (bottom to top) logarithmically spaced between $10$ and $10^{6}$~\flux. The dotted lines show the $1/M_p$ scaling expected from ``energy'' limited flows. The dashed line shows the predicted transition between recombination limited flows and energy limited flows from Section~2, above the line is predcited to be recombination limited and below is predicted to be energy limited.}\label{fig:mdot_mass}
\end{figure}

We can see this transition in more detail if we plot the mass-loss rate as a function of flux for various planet masses, for the full simulated range of incident flux. The results are shown in Figure~\ref{fig:mdot_mass} with the $1/M_p$ scaling for energy-limited flows shown as dotted lines. We find at low fluxes the simulation is well matched by an energy limited scaling; however, at higher fluxes the trend becomes much flatter indicating recombination limited flows (Noting that in this parameter range the planet's escape temperature is $<10^{4}$~K so mass plays a limited role as discussed above). The dashed line in Figure~\ref{fig:mdot_mass} shows the predicted analytic transition from recombination to energy limited flows. Above the line is predicted to be recombination limited and below is predicted to be energy limited. We see this prediction agrees well with the numerical results. 

\begin{figure}
\centering
\includegraphics[width=\columnwidth]{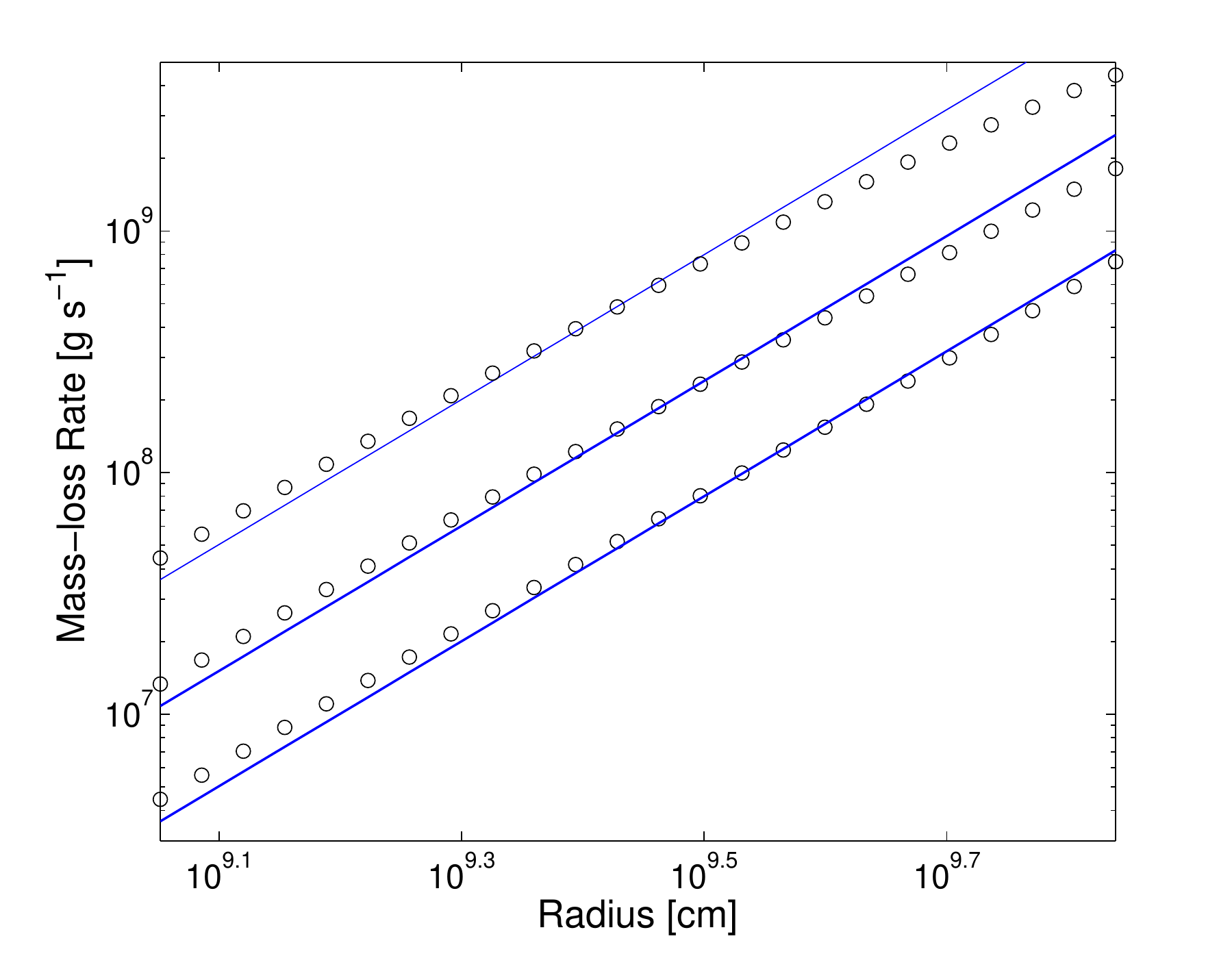}
\caption{Mass-loss rate as a function of planet radius. Simulation results are shown as points for fluxes (bottom to top) of $\sim$ 10, 30, 100 \flux. The solid lines shows the ``energy'' limited mass-loss rate with an efficiency of $ 20\%$.}\label{fig:mdot_rad}
\end{figure}

Finally, in order to measure the efficiency of the flow we measure the mass-loss rate as a function of planet radius (due to its steep dependence this should provide a good range of mass-loss rates over which to tune this parameter). The discussion in Section~2 indicates that at low fluxes and low masses the flow should be strongly energy limited, which is indeed what we find. In Figure~\ref{fig:mdot_rad} we show the mass-loss rate as a function of radius for a planet of mass 10~M$_\oplus$ and fluxes of $\sim 10$-100 (points). We find that an mass-loss rate with an energy efficiency of $20$\% agrees well with the simulations\footnote{Note in order to compare with Equation~\ref{eqn:EL}, we must drop the factor 4 in the denominator that arises from geometrical averaging which does not occur in spherically symmetric simulations.} and as such represents a suitable measure of the mass-loss efficiency in the region of parameter space.  

\section{Discussion}\label{sec:discussion}

We have argued that evaporation due to UV radiation can exist in three limiting cases: ``recombination-limited'' - where the flow is in radiative-recombination equilibrium and approximately isothermal at $10^4$~K; ``energy-limited'' where incoming radiation is balanced by mechanical luminosity; and a new regime:``photon-limited'' - where the mass-loss is limited by the fact that each ionizing photon can allow one hydrogen atom to escape.  Each distinctive limit has a different scaling with planet mass, radius as well as the incident flux. We have shown that there is no unique transition flux from one regime to the other and such a limit depends on the planetary parameters. 

Schematically if the flux is high enough such the the density in the flow is large (and correspondingly the recombination time is short) or the planet is so big or massive that the flow time-scale is long then the problem is in the radiative-recombination limited, where the recombination time is much faster than the flow-time-scale. As argued by Murray-Clay et al. (2009); Owen \& Jackson (2012) in such a scenario it behaves as a HII region that is approximately isothermal  and the density at the ionization front scales as $J_0^{1/2}$, in this region the mass-loss rate (in a 1D calculation) is analytically tractable and we present such a calculation in Section~ 2.1.2. 

If one decreases the flux, the density in the ionization front will drop and the recombination time will increase. Once it reaches the point where the recombination time is longer than the flow time-scale then recombinations can be ignored.  When the recombination time is much longer than the flow time then we also find the cooling time is longer than the recombination time and subsequently the flow time-scale. In this case radiative cooling can be ignored. This is what one thinks off as standard ``energy-limited'' flow. However, we note that when the recombination time is of order the flow-time scale then the cooling time is shorter than the recombination time and thus shorter than the flow time-scale. Thus, one expects the transition from recombination to energy limited to happen smoothly and broadly over a large range of fluxes. This is indeed what our numerical calculations indicate, that the transition occurs over several orders of magnitude in flux. This is different to the case of X-ray heating discussed by Owen \& Jackson (2012), where since the heating and cooling times scaled identically with flux then the cooling time was always comparable or shorter than the flow time-scale. Thus, in the case of X-ray heating Owen \& Jackson (2012) argued that the flow was never ``energy-limited''.  

If when one reaches the point where the recombination time is becoming longer than the flow time-scale and the planet is not tightly bound (such that the energy limit is not restrictive) then the mass-loss rate will be limited by the number of ionizing photons. Such that one can only lose one hydrogen atom per ionizing photon. Obviously, in this limit planet mass is completely irrelevant and the mass-loss rate just scales as the flux $\times$ area.  Such a case is similar to the evaporation of low-mass neutral clumps by nearby young massive stars discussed by Bertoldi \& Mckee (1990). However, this regime will only be restricted to a small (but perhaps important) regime of parameter space for low-mass large radii planets, similar to the masses and radii of the {\it Kepler} planets at young 10s Myr ages (e.g. Lopez et al. 2012, Owen \& Wu, 2013). Unfortunately, such a problem could not be investigated with realistic planet parameters in 1D spherically symmetric simulations. This is because such simulations assume spherical irradiation, in this case the absorbing area would be $4\pi R_{\rm IF}^2$. However, in the reality of planet-parallel irradiation and the fact the ionization front is marginally optically thin then we suspect the absorbing area is more likely $\sim\pi R_p^2$. In this regime the ionization front is thick and by construction has a radius $R_{\rm IF}\gg R_p$ and as such the ``photon-limited'' mass-loss rate from the simulations is much higher than reality and any other limit. One needs to perform 2/3D simulations to investigate the role of ``photon-limited'' mass-loss in planetary mass-loss. 

Our numerical calculations indicate that the hydrodynamic efficiency if fairly low for low-mass planets $\sim 10-20$~\%, comparable to the values used in planetary evolution studies (e.g. Lopez et al. 2012, Lopez \& Fortney 2013) indicating that much of the absorbed energy remains as kinetic/thermal energy in the flow rather than being converted to mass-loss. When the flow transitions to energy or photon limited the ionization front is thick; essentially encompassing the entire the flow. Furthermore, we note that not all the heating is concentrated at the planet's surface as can be seen in the bottom panels of Figure~\ref{fig:topology}, where the maximum temperature is reached at $\sim 2$ planetary radii. The ionization fraction increases monotonically with radius but still contains a significant neutral fraction out to many planetary radii. Therefore simplified calculations assuming infinitesimally thin ionization fronts are not appropriate in this limit. This means photon hardening can play an important role in pre-heating material; {\bc this occurs when harder photons penetrate deeper than the standard ionizing ($\sim 13.6$~eV) photons resulting in an average photon energy that is harder than the total spectrum, these photons 'pre-heat' the underlying gas}. Thus, one requires a non-monchromatic numerical method to take this into account (unlike Murray-Clay et al. 2009, who assume a spectrum with just 20~eV photons). 

Finally, we note we have presented a new numerical algorithm for the inclusion of UV heating in planetary evaporation simulations. We have included the frequency dependence of the ionizing spectrum for pure Hydrogen atmospheres and note this can easily be extended to include more species (Friedrich et al. 2012). Such an algorithm can be incorporated into multi-dimensional calculations that include ray-tracing (e.g. Owen \& Adams, 2014) to investigate planetary evaporation in 2/3D. Such a further step is important, not only to investigate the flow properties, in particular ``photon-limited'' evaporation identified here; but, also to produce realistic calculations that can be compared to observational indicators of planetary evaporation. This includes measured transits in Lyman-$\alpha$ (Vidal-Madjar et al. 2003, Ben-Jaffel et al. 2007, 2008, Lecavelier Des Etangs et al. 2010) and other metal lines (Vidal-Madjar et al. 2004, Ballester \& Ben-Jaffel 2015).

\section{Summary}
We have investigate planetary evaporation under the influence of irradiation by ionizing UV photons at close enough separations such that evaporation takes place in the hydrodynamic limit. We argue that there are three main evaporation limits. The first two were demonstrated by Murray-Clay et al. (2009), at high fluxes the flow is in radiative-recombination equilibrium and the mass-loss rate scales as the square-root of the ionizing flux; the second where recombinations are negligible and the mass-loss rate is limited by the amount of incoming energy and as such scales linearly with the ionizing flux. Finally, we argue that there is a third limiting factor that must be taken into account, which we term ``photon-limited'' mass-loss. In photon-limited mass-loss one can only lose a single hydrogen atom per incoming ionizing photon and is important when the planets gravity is weak (low density, low mass planets). 

We present a basic physical picture of the three different evaporation regimes and show that transitions from one regime to another does not happen a specific UV flux, as assumed in some recent evolutionary studies (e.g. Jin et al. 2014). The different evaporation regimes dominate different planetary parameters and different flux and we present a framework that can be incorporated in to planet evolution studies, provided a mass-loss efficiency is chosen for the energy-limited flows.

We have presented a new numerical algorithm for performing radiation-hydrodynamic simulations that takes into account the frequency dependence of the irradiation spectrum. It does not require the ionization front to be thin, or approximately isothermal and self consistently finds the correct ionization and thermodynamic structure of the flow. This radiative transfer scheme has been incorporated into the {\sc zeus} hydrodynamic code. 

Using 1D spherically symmetric numerical simulations we have investigated the transition between ``energy-limited'' flows and ``recombination-limited'' flows. We take into account the non-monochromatic nature of the ionizing photons unlike previous works (e.g. Murray-Clay et al. 2009). The transition from recombination limited to energy limited occurs when the recombination time becomes longer than the flow time-scale and the ionization front encompasses the entire flow scale. In this case radiation hardening from the full spectrum is an important contribution to the thermodynamics. 

We find that the transition from recombination limited to energy limited is broad and takes several orders of magnitude in flux. Planets with deeper potentials transition to energy-limited flows at higher fluxes consistent with our analytically expectations. Furthermore, we find that in the`energy-limited case efficiencies around 10-20\% provide a good fit to mass-loss rates.

Finally, we could not use our 1D spherically symmetric simulations to investigate ``photon-limited'' flows and future mutli-dimensional calculations are required to investigate the importance of this type of flow. Something that is possible with our presented algorithm and we plan to incorporate this into the multi-dimensional ray-tracing scheme for planetary evaporation presented by Owen \& Adams, (2014).

\acknowledgments
We thank the two anonymous referees for constructive reports that improved the article. We are grateful to Noam Soker and Ealeal Bear for comments on the original manuscript. JEO acknowledges support by NASA through Hubble Fellowship grant HST-HF2-51346.001-A awarded by the Space Telescope Science Institute, which is operated by the Association of Universities for Research in Astronomy, Inc., for NASA, under contract NAS 5-26555. JEO is grateful to CITA for hospitality during the completion of this work. Several of the numerical calculations were performed on the Sunnyvale cluster at CITA, which is funded by the Canada Foundation
for Innovation. 


\end{document}